\documentclass[aps,prletter,print,superscriptaddress]{revtex4-2}
\usepackage{graphicx}
\usepackage{tikz}
\usepackage{mathrsfs}
\usepackage{todonotes,comment,lineno}
\usepackage{amsmath,amsfonts}
\usepackage[font=small,skip=0pt,justification=raggedright]{subcaption}
\usepackage[font=small,skip=0pt,justification=raggedright]{caption}
\usepackage{todonotes}
\usepackage{siunitx}
\usepackage[normalem]{ulem}
\usepackage{soul}

\usepackage{hyperref}

\newcommand{\bk}{\boldsymbol{\kappa}}
\newcommand{\bU}{\mathbf{U}}

\def\tu{{\overline u}}

\def\bu{{\bf u}}

\def\cC{\mathcal{C}}

\pgfmathdeclarefunction{gauss}{2}{
    \pgfmathparse{1/(#2*sqrt(2*pi))*exp(-\left( x-#1)^2)/(2*#2^2))}
}

\newcommand{\solidrule}[1][1cm]{\rule[0.5ex]{#1}{1pt}}

\definecolor{mplblue}{RGB}{55,126,184}
\definecolor{mplred}{RGB}{228,26,28}
\definecolor{mplgreen}{RGB}{77,175,74}
\definecolor{mplorange}{RGB}{255,127,0}
\definecolor{mplpurple}{RGB}{152,78,163}
\definecolor{mplbrown}{RGB}{166,86,40}
\definecolor{mplblack}{RGB}{0,0,0}
\definecolor{mplwhite}{RGB}{255,255,255}

\definecolor{mpl1}{RGB}{102,194,165}
\definecolor{mpl2}{RGB}{190,186,218}
\definecolor{mpl3}{RGB}{231,138,195}
\definecolor{mpl4}{RGB}{140,81,10}

 \newcommand\gryst{\bgroup\markoverwith{\textcolor{blue}{\rule[0.5ex]{2pt}{0.4pt}}}\ULon}
\def\gry#1{{#1}}
\def\rdm#1{{#1}}

\begin{document}

\title{
Numerical dispersion effects on the energy cascade in large-eddy
simulation 
}

\author{Gopal R. Yalla}
\affiliation{The Oden Institute for Computational Engineering and Sciences, The University of Texas at Austin}
\author{Todd A. Oliver}
\affiliation{The Oden Institute for Computational Engineering and Sciences, The University of Texas at Austin}

\author{Robert D. Moser}
\affiliation{The Oden Institute for Computational Engineering and Sciences, The University of Texas at Austin}
\affiliation{Department of Mechanical Engineering, The University of Texas at Austin}

\date{\today}

\begin{abstract}
Implicitly filtered large-eddy simulation (LES) is by nature
numerically under-resolved.  With the sole exception of
Fourier-spectral methods, discrete numerical derivative operators
cannot accurately represent the dynamics of all of the represented
scales. Since the resolution scale in an LES usually lies in the
inertial range, these poorly represented scales are dynamically
significant and errors in their dynamics can affect all resolved
scales. This Letter is focused on
characterizing the effects of numerical dispersion error by studying
the energy cascade in LES of convecting homogeneous isotropic
turbulence.  Numerical energy and transfer spectra reveal that
energy is not transferred at the appropriate rate to wavemodes where
significant dispersion error is present. This leads to a deficiency of
energy in highly dispersive modes and an accompanying pile up of
energy in the well resolved modes, since dissipation by the subgrid
model is diminished. An asymptotic analysis indicates that dispersion
error causes a phase decoherence between triad interacting wavemodes,
leading to a reduction in the mean energy transfer rate for these
scales. 
These findings are relevant to a wide range of LES, since turbulence commonly
convects through the grid in practical simulations. Further, these results
indicate that the resolved scales should be defined \rdm{to not include the
dispersive modes}.
\end{abstract}

\maketitle
In numerical analysis, one generally aspires to make the resolution sufficiently
fine so that discretization error is negligible. However, large-eddy simulation
(LES) is, by definition, under-resolved. In practice, information about the
small-scale turbulence is discarded by numerical
discretization. This includes the projection of the \gry{infinite-dimensional}
velocity field onto a \gry{finite-dimensional} solution space (often referred to as
the implicit filter) and the introduction of 
numerical derivative operators, which together characterize the scales in the
resolved field whose dynamics are accurately represented.  Since the resolution
scale in an LES often lies in the \gry{energy-containing} inertial range, the effects of
discretization error must generally be considered, with the sole
exception being for Fourier-spectral methods. 
In this Letter, we examine the statistical consequences of
this dispserion error in LES.

A simple flow in which to explore these dispersion error effects is
infinite Reynolds number forced homogeneous isotropic turbulence with
fluctuating velocities $\bu$ transported with a uniform convection
velocity $\bU$, with magnitude $U$. We consider an LES with homogeneous isotropic
resolution $\Delta$, and following \citet{moser2020statistical} write
the filtered incompressible Navier-Stokes equations for the
fluctuating velocity for this case as
\begin{align}
\frac{\partial \tu_i}{\partial
t}+U_j\frac{\delta \tu_i}{\delta x_j} + \frac{\delta\mathcal{F}({\tu_i\tu_j})}{\delta x_j} &
=-\frac{\delta \bar{p}}{\delta x_i}
+ \frac{\delta \tau_{ij}}{\delta x_j}
-\cC_i^m + f_i
\label{eq:filtNS2}\\
\frac{\delta\tu_i}{\delta x_i}  &= -\cC^c\label{eq:filtCont2}
.
\end{align}
where $p$ is the pressure, $f_i$ is the forcing, $\overline{\cdot}$ designates a filtered
quantity, $\mathcal{F}(\cdot)$ is a (potentially different) filtering
operator that maps the nonlinear terms to the LES solution space, and
$\tau_{ij} = - \overline{u_i u_j}
+ \mathcal{F}({\overline{u}_i \overline{u}_j})$ is the subgrid
stress. Furthermore, $\delta/\delta \mathbf{x}$ denotes the discrete
derivative operator. The terms $\cC_i^m$ and $\cC^c$ are generalized
commutators \citep{yalla2021effects} that, in this case, solely account for the discretization errors in the
numerical derivatives. More generally, they represent the difference
between the exact terms in the filtered equations and the approximate
terms used in the LES computations. The approximations to the pressure
gradient and continuity equation are not of interest here. Instead we focus on
the linear and nonlinear convective terms and their contributions
$\cC_i^l$ and $\cC_i^n$, respectively, to the generalized commutator $\cC_i^m$, where:
\begin{equation}
\cC_i^l=U_j\left(\overline{\frac{\partial u_i}{\partial x_j}}
- \frac{\delta \tu_i}{\delta
  x_j}\right) \qquad \cC_i^n=\overline{\frac{\partial\mathcal{F}({\tu_i\tu_j})}{\partial
  x_j}}-\overline{\frac{\partial \tau_{ij}}{\partial x_j}}
  - \frac{\delta\mathcal{F}({\tu_i\tu_j})}{\delta x_j}
  + \frac{\delta \tau_{ij}}{\delta x_j}
  .
\end{equation}
Usually, these commutators are neglected in an LES.

To assess the consequences of neglecting $\cC_i^m$, we solve
(\ref{eq:filtNS2}) and (\ref{eq:filtCont2}) assuming $\cC_i^m=0$ for
forced homogeneous isotropic turbulence with periodic boundary
conditions.  Both filters $\overline{\cdot}$ and
$\mathcal{F}({\cdot})$ are defined to be Fourier cutoff filters, to
allow us to isolate dispersion effects from aliasing
effects. However, simulations using a collocation
projection for $\mathcal{F}$ yielded results similar to those
presented here.

\begin{figure}[t!]
	\centering
	\includegraphics[width=0.5\textwidth]{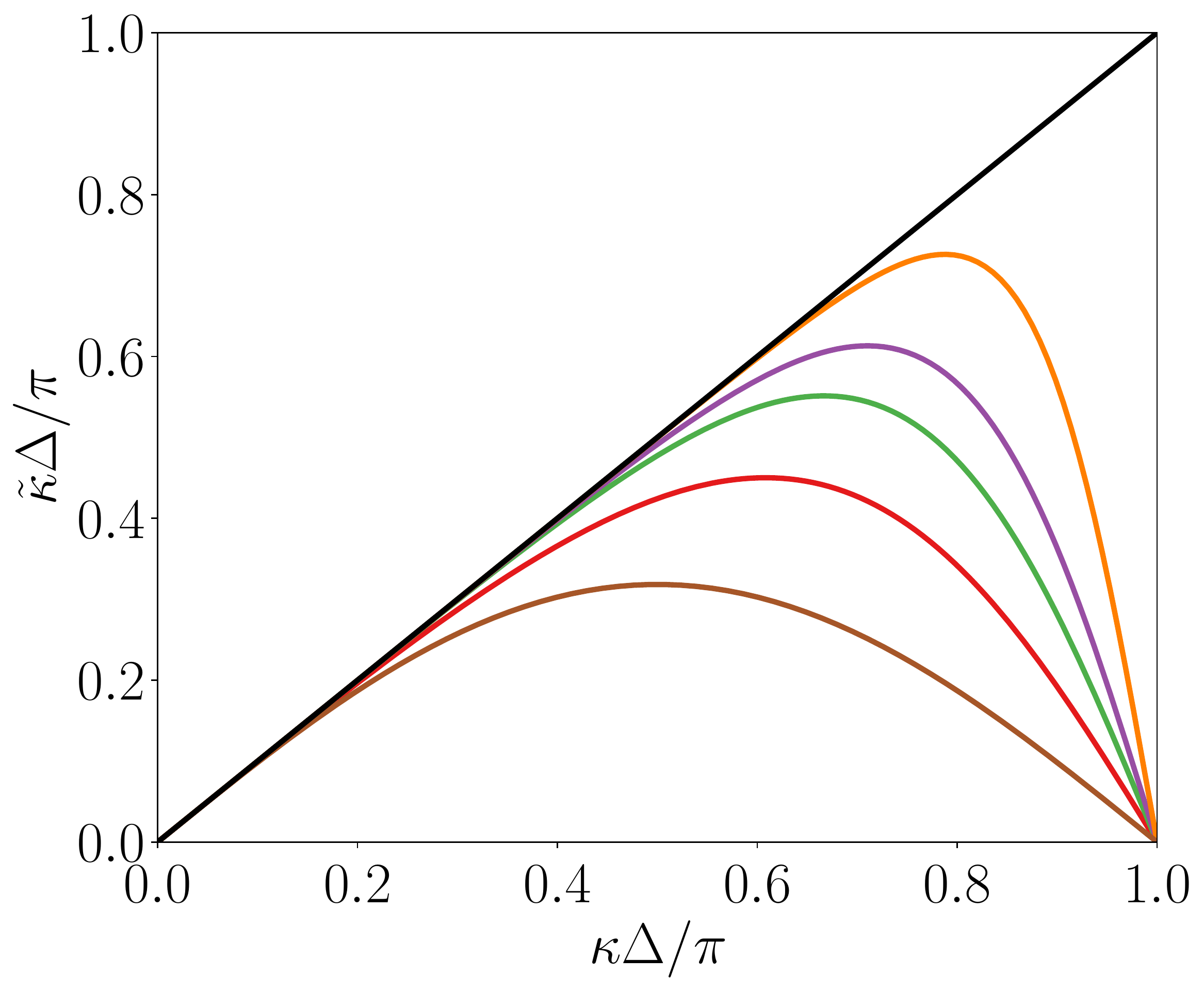}
	\caption{ The effective wavenumbers corresponding to the first derivative operators
		for a Fourier-spectral method (\protect\solidrule{}), 7th order B-spline
		collocation method (\textcolor{mplorange}{\solidrule{}}), 4th order
		B-spline collocation method (\textcolor{mplpurple}{\solidrule{}}), 3rd order
		B-spline collocation method (\textcolor{mplgreen}{\solidrule{}}),
		2nd order B-spline collocation method (\textcolor{mplred}{\solidrule{}}), 
	and \gry{2nd order centered difference method} (\textcolor{mplbrown}{\solidrule{}}).}
	\label{fig:effective_wavenumbers}
\end{figure}

\gry{Six} different discrete first derivative
operators $\delta/\delta \mathbf{x}$ are used: a Fourier-spectral method, \gry{a 2nd
order centered difference method}, and 2nd, 3rd, 4th, and 7th order B-spline
collocation methods.
All the operators are implemented in a modified version of the Fourier spectral code
{\rm \small POONGBACK} \cite{lee2015direct} by substituting the effective wavenumbers associated
with each method for the true wavenumber when evaluating first derivatives
in Fourier-space. Specifically, let $i\tilde{\kappa}(\kappa)$ be the eigenvalue of the first
numerical derivative operator that corresponds to
the eigenfunction $e^{i\kappa x}$. Then $\tilde\kappa$
is the effective wavenumber for the first
derivative operator, which is plotted in 
Figure~\ref{fig:effective_wavenumbers} for the numerical derivative
operators of interest here. The effective wavenumber for the first
derivative in the $i^{\rm th}$ coordinate direction is then
$\tilde\kappa_i=\tilde\kappa(\kappa_i)$ and an effective wave
``vector'' $\tilde{\bk}$ can be defined with components $\tilde\kappa_i$.
To isolate the effects of numerical dispersion, the Laplacian 
that arises from the eddy viscosity model for the subgrid stress because
$\nu_t$ is a constant (see below) is treated spectrally in all the
LES performed here.

A third-order low storage Runge-Kutta method \citep{spalart1991spectral} is used for time advancement,
\rdm{with time step selected to maintain CFL=0.5 as defined
in \citet{spalart1991spectral}}. 
The filtered Navier-Stokes equations are solved using the
vorticity-velocity formulation of \citet{kim1987turbulence}; see
Appendix~\ref{sec:A1} for more details regarding the implementation of the 
vorticity-velocity formulation for nonspectral numerics.
The skew-symmetric form of the nonlinear terms is used in the LES
because it
conserves energy and has been reported to perform
well in the presence of aliasing and discretization errors in LES
\cite{zang1991rotation,tadmor1984skew,chow2003further,blaisdell1996effect}.
In addition, the consequences of dispersion error in other forms of the
nonlinear terms are discussed.

The forcing $f_i$ is formulated to inject energy at a constant rate
$\varepsilon$ as in \citep{mohan2017scaling} and the simulations are performed in a
cubical domain of size $L$. Because the turbulence is
statistically stationary, $\varepsilon$ is also the \gry{mean} rate of kinetic
energy dissipation. Unless otherwise indicated, all quantities are
reported in units in which $\varepsilon=1$ and $L=2\pi$. The filtered
velocity is represented with \gry{32} Fourier modes in each direction, for
an effective uniform resolution of \gry{$\Delta=2\pi/32$}, and the forcing $f_i$
is active only in the wavenumber range $0<|\bk|\le 2$. 
\gry{The filter $\mathcal{F}({\cdot})$ (a Fourier cutoff) is applied to the
	nonlinear terms as in a dealiased pseudo-spectral method, by
	evaluating the nonlinear product on a $48^3$ grid and truncating
	the discrete Fourier transform of the result to $32^3$ Fourier
	modes (the 3/2 rule) \citep{orszag1971elimination}.}
The subgrid stress is approximated by a
Kolmogorov model $\tau_{ij} = 2\nu_tS_{ij}$, where $\nu_t= C_m \Delta^{4/3}\varepsilon^{1/3}$
is the eddy viscosity, $S_{ij} = \frac{1}{2}(\delta_i \bar{u}_j + \delta_j \bar{u}_i)$ is the filtered strain rate
tensor and $C_m = 0.065$ \citep{leslie1979application,haering2019resolution}.
Note that the phenomenon reported here is not dependent on this choice
of the subgrid model.

Letting $\hat\cdot$ denote the Fourier transform and $\cdot^*$ denote the
complex conjugate, the evolution equation for the instantaneous
resolved energy
spectrum $E(\bk,t) = \frac{1}{2}
\hat{\bar{u}}_j^*(\bk,t) \hat{\bar{u}}_j(\bk,t)$ in the LES is
\begin{equation}
	\frac{\partial E(\bk,t)}{\partial t} = T_{\rm N}(\bk,t) - 2\nu_t
	|\bk|^2 E(\bk,t) + F(\bk,t)
	\label{eq:inst_energy}
	,
\end{equation}
where $T_{\rm N}$ is the numerical transfer spectrum that represents
the exchange of energy between wavenumbers due to triad interactions,
in the presence of numerical dispersion error, and
$F={{\rm Re}}\left\{ \hat{\bar u}^*_j(\bk,t)\hat f_j(\bk,t) \right\}$
is the spectrum of the energy
production arising from the forcing. For the skew-symmetric
form of the nonlinear terms used here, $T_{\rm N}=T_{\rm skew}$ with  
\begin{equation}
	T_{\rm skew}(\bk,t) = -\frac{1}{2}{\rm Im} \left\{ \sum_{\bk'} 
		(\tilde{\kappa'_\ell} + \tilde{\kappa}_\ell) \hat{\bar{u}}_k(\bk,t) \hat{\bar{u}}^*_k(\bk',t) \hat{\bar{u}}^*_\ell
	(\bk-\bk',t)  \right\}
	.
	\label{eq:Tskew}
\end{equation}
See Appendix~\ref{sec:A2} for further discussion of the transfer spectrum
for this and other nonlinear forms. 
The LES results reported here are the
one-dimensional energy spectra in the direction of the convection
\rdm{velocity, which is chosen to be aligned with the grid direction}
($i=1$). The energy spectra are computed
as ${E}_{\rm 1D}(\kappa_1) = \sum_{\kappa_2,\kappa_3}\langle E(\bk,t)\rangle$,
where $\langle\cdot\rangle$ indicates the expected value, which is
approximated as a time average. In
addition, analogous one-dimensional transfer
spectra $T_{\rm 1D}(\kappa_1)=\sum_{\kappa_2,\kappa_3}\langle
T_{\rm skew}(\bk,t)\rangle$ are reported. \rdm{When the convection velocity
  is not aligned with the grid, dispersive effects on the
one-dimensional spectra similar to those
reported here occur in each grid direction in which $\bU$ has
a nonzero component.}

\begin{figure*}[t!]
\begin{subfigure}[b]{0.47\textwidth}
\vskip\baselineskip
\centering
\includegraphics[width=1\textwidth]{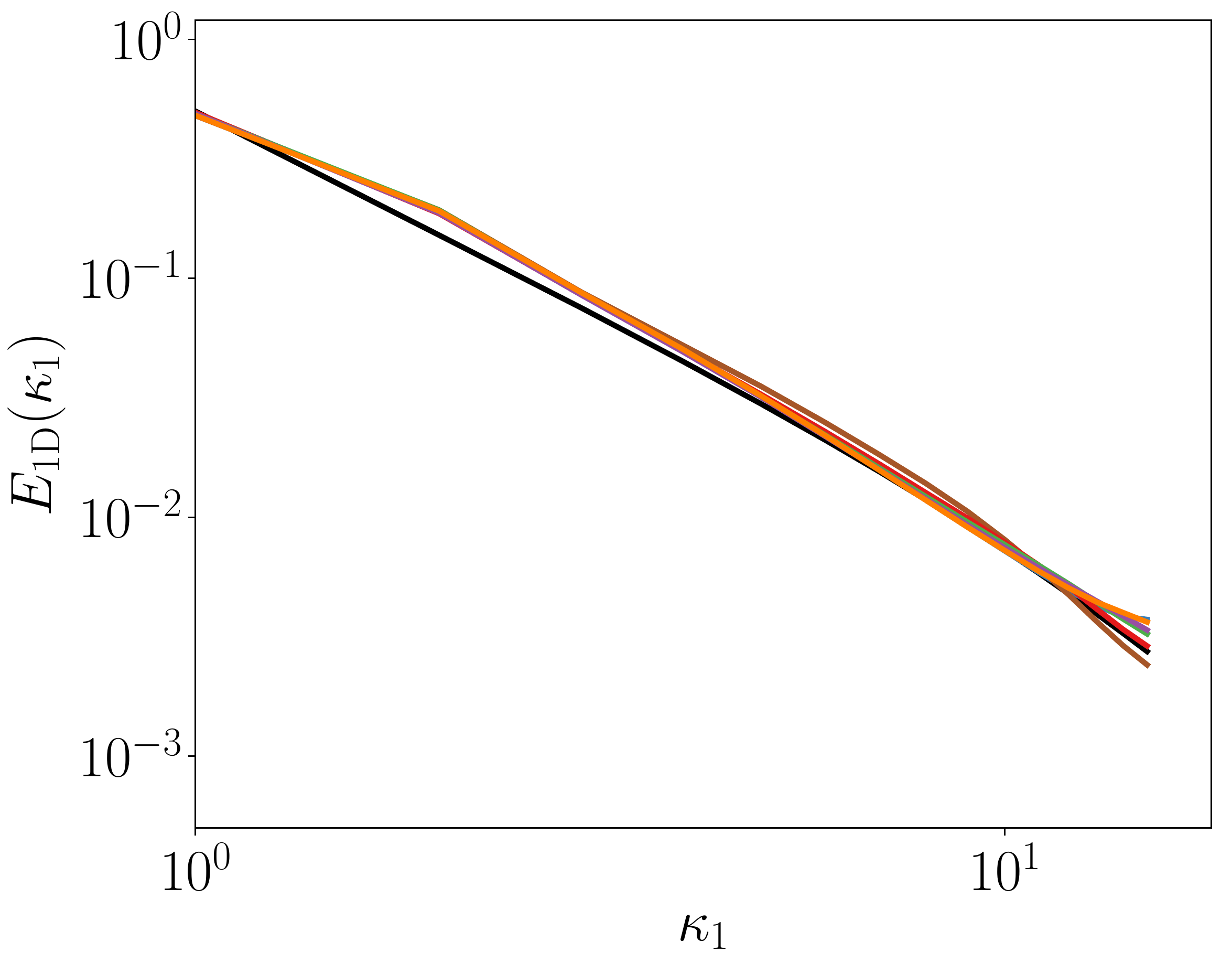}
\caption{$E_{1D}$ for $U=0$}
\label{fig:E0}
\end{subfigure}
\hspace{1em}
\begin{subfigure}[b]{0.47\textwidth}
\centering
\includegraphics[width=1\textwidth]{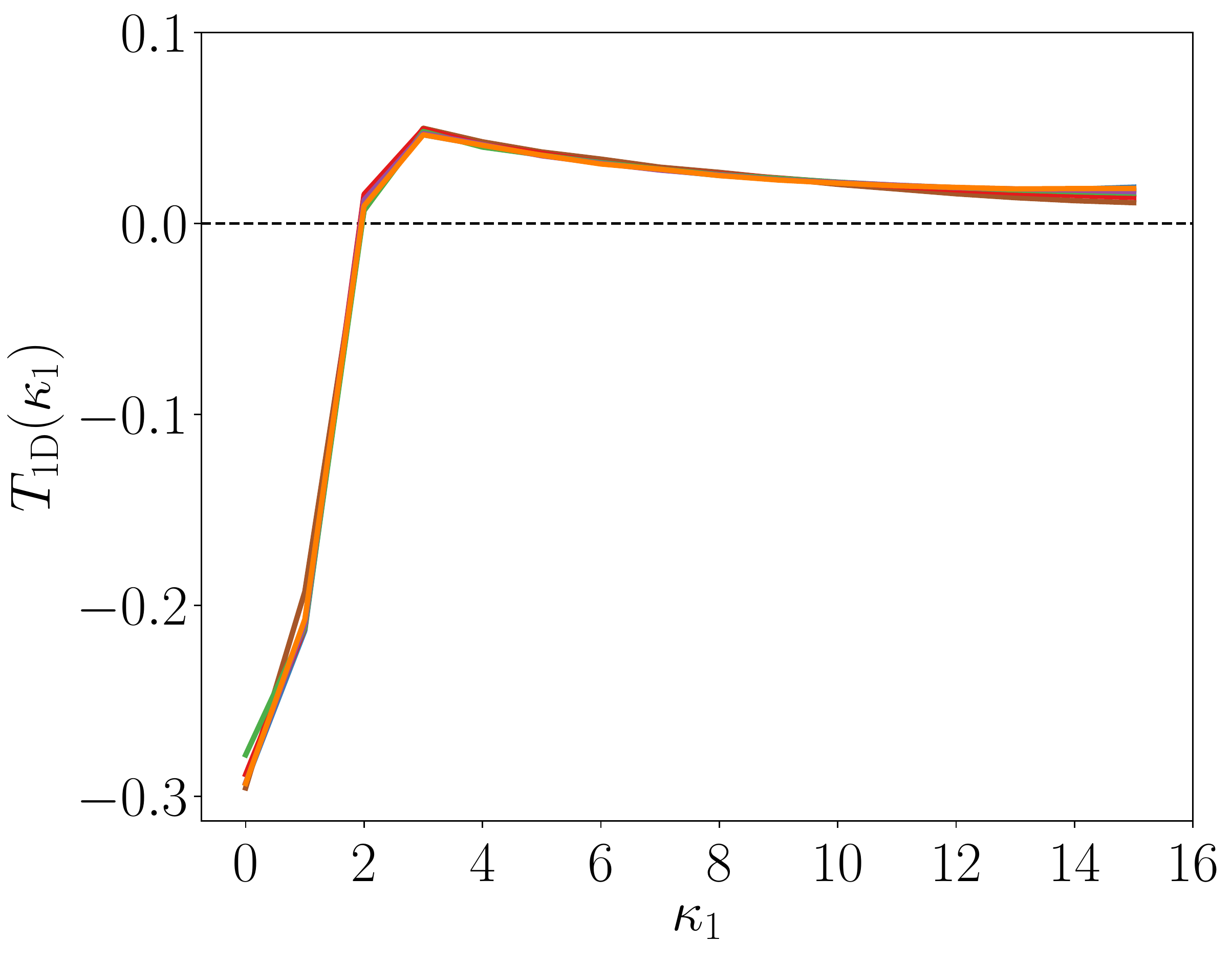}
\caption{$T_{1D}$ for $U=0$} 
\label{fig:T0}
\end{subfigure}
\vskip\baselineskip
\begin{subfigure}[b]{0.47\textwidth}
\centering
\includegraphics[width=1\textwidth]{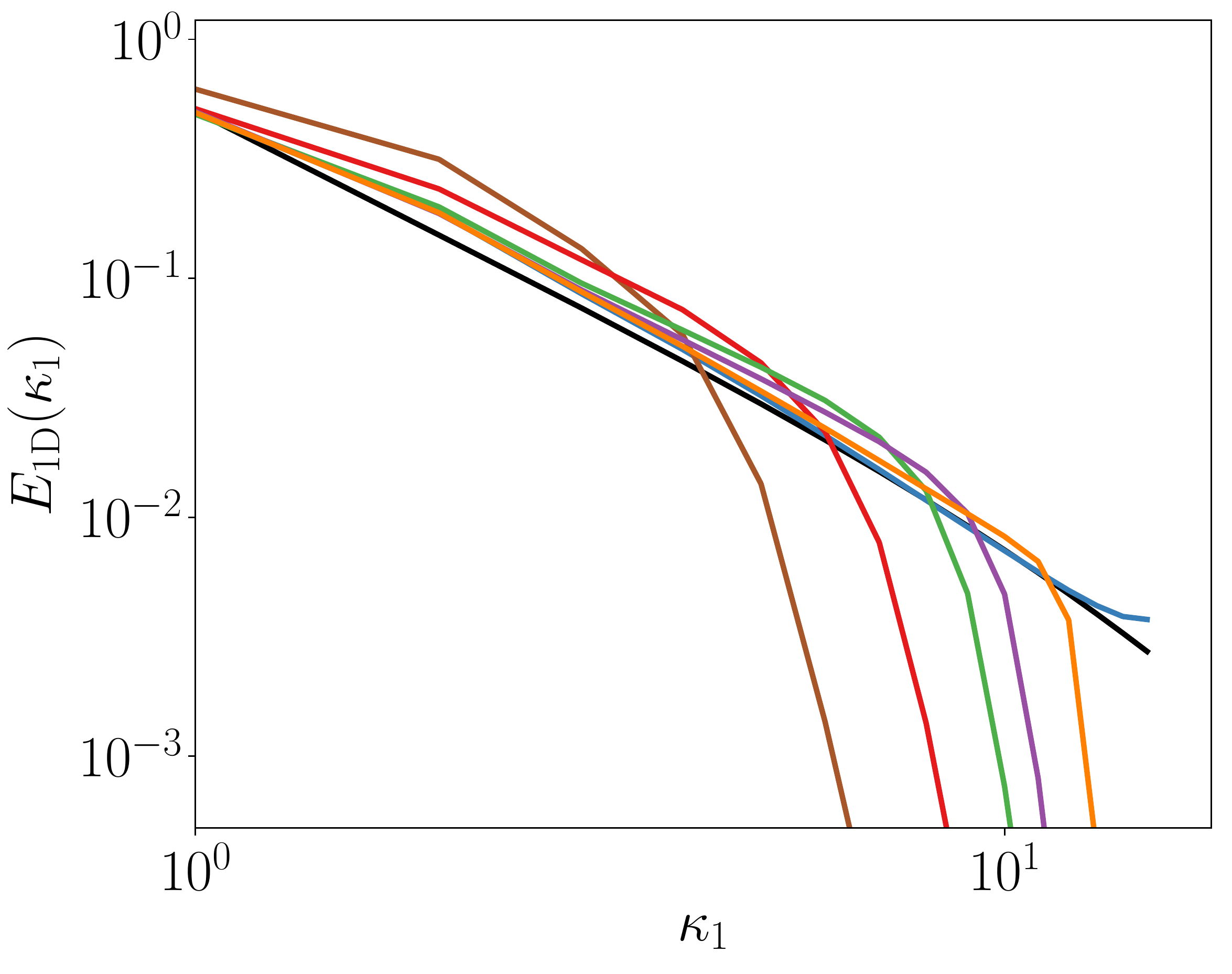}
\caption{$E_{1D}$ for $U=35$}
\label{fig:E50}
\end{subfigure}
\hspace{1em}
\begin{subfigure}[b]{0.47\textwidth}
\centering
\includegraphics[width=1\textwidth]{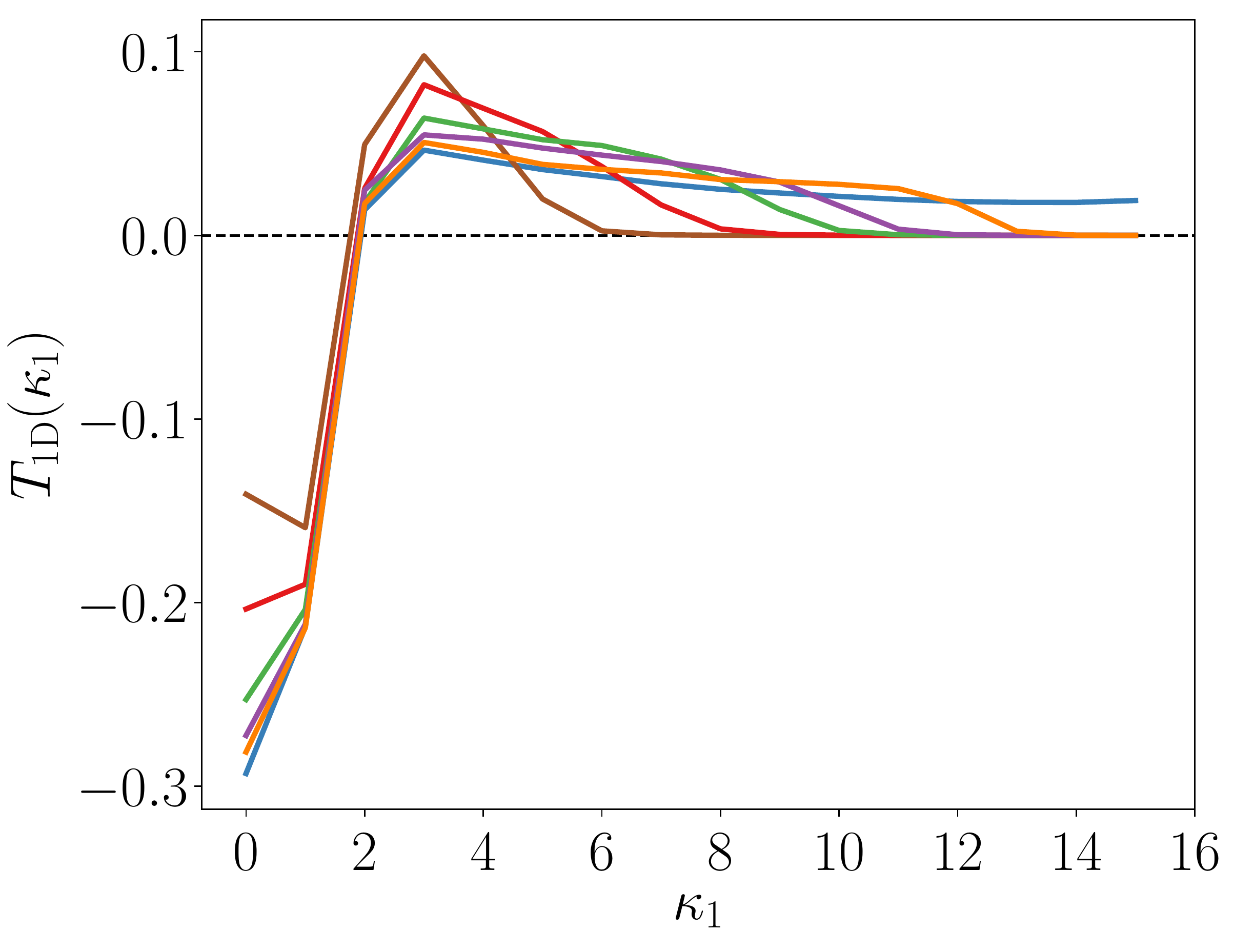}
\caption{$T_{1D}$ for $U=35$} 
\label{fig:T50}
\end{subfigure}
\caption{One dimensional spectra in the convection direction 
  of energy $E_{\rm 1D}$ and energy transfer rate $T_{\rm 1D}$, with convection
  velocities $U=0$ and $35$.
Shown are spectra from
theory (\protect\solidrule{}), 
Fourier-spectral method (\textcolor{mplblue}{\solidrule{}}), 
7th order B-spline collocation method (\textcolor{mplorange}{\solidrule{}}), 
4th order B-spline collocation method (\textcolor{mplpurple}{\solidrule{}}),  
3rd order B-spline collocation method (\textcolor{mplgreen}{\solidrule{}}),
2nd order B-spline collocation method (\textcolor{mplred}{\solidrule{}}),
\gry{2nd order centered difference method} (\textcolor{mplbrown}{\solidrule{}}).
}
\label{fig:results}
\end{figure*}

In the absence of mean convection ($U=0$), each numerical scheme
produces spectra that agree with an equivalently filtered
$\kappa^{-5/3}$ theoretical inertial range spectrum (see
Figure~\ref{fig:E0}). Furthermore, the transfer spectra are identical
to that for the
Fourier-spectral case for all the numerical approximations (see Figure~\ref{fig:T0}).
For reference, the statistical characteristics of the turbulence for the
$U=0$, Fourier-spectral case are reported in Table~\ref{tab:stats}.

To demonstrate numerical dispersion effects, consider the case
with mean convection velocity
\rdm{$U=35$}. In this case, the value of \rdm{$U/u'\approx 27$} ($u'$ is the
root-mean square velocity as defined in Table~\ref{tab:stats}) 
is comparable to that at the centerline of a turbulent channel flow, where $U/u'$
ranges from 23 to 30 for friction Reynolds number ranging from 180 to
5200 \cite{lee2015direct}.
As expected, for spectral numerics, no change
from the $U=0$ case occurs in either the energy spectrum or transfer
spectrum. However, for all other numerical schemes, the
one-dimensional energy spectra in the direction of convection are
significantly reduced over a range of the highest resolved wavenumbers
(see Figure~\ref{fig:E50}).  The corresponding transfer spectra in the
direction of convection tend to zero over this range of resolved modes
(see Figure~\ref{fig:T50}). 
In effect, numerical dispersion error
prevents energy from transferring at the appropriate rate from the
largest to smallest resolved scales. As a consequence, energy also
piles up in the larger resolved scales as energy is not transferred to
the smallest resolved scales at the correct rate for the subgrid
model to dissipate (see Figure~\ref{fig:E50}).  For energy conserving
numerics, we must have $\sum_{\bk} T_{\rm N}(\bk,t)= 0$ for all
$t$. To maintain this balance, the energy transfer spectrum at
all wavenumbers is affected by the dispersion error that is primarily
in the largest wavenumbers (see Figure~\ref{fig:T50}).
Moreover, the energy transfer rates in directions orthogonal to convection are
not degraded, however, the energy spectra in the orthogonal directions are
impacted by the errors in the convection direction (not shown).

\def\arraystretch{1.25}
\begin{table}[t!]
\begin{center}
	\captionsetup{justification=centering}
	\caption{Statistical characteristics of LES turbulence averaged over 500
			eddy turnover times\\ for the $U=0$ case with Fourier spectral
	numerics. Values are normalized by $\varepsilon$ and $L/2\pi$.}
	\begin{tabular}{@{\hspace{1em}}l@{\hspace{1em}}@{\hspace{1em}}c@{\hspace{1em}}}
    \hline
    \hline
	Resolved kinetic energy, $k_{res} = \frac{1}{2} \langle \bar{u}_i \bar{u}_i \rangle$ &
	2.467 \\
	\gry{RMS velocity, $u' = \sqrt{(2/3)k_{res}}$} & 1.282 \\
	Integral scale, $\mathcal{L} = \frac{\pi E_{\rm 1D}(0)}{{u'}^2}$ & 1.129 \\
	Large eddy turnover time, $T_{L} = \mathcal{L}/u'$ & 0.880 \\
    \hline
    \hline
    \label{tab:stats}
\end{tabular}
\end{center}

\end{table}

The reason for the observed degradation of energy transfer to the
smallest resolved scales when $U>0$ can be understood through analysis
of a case in which $U\gg u'$. 
\gry{Let $\epsilon=u'/U\ll1$ be a small parameter (not to be confused with the
mean dissipation rate $\varepsilon$)}. Then the velocity Fourier
coefficients $\hat{\bar u}_i$ vary on a fast and a slow time
scale. Using a multiscale asymptotic representation, $\hat{\bar u}_i$
can be said to depend on a fast time variable $t_f=t/\epsilon$ and a
slow time variable $t_s=t$. Further, as with Taylor's hypothesis, in
the continuous case, $\partial/\partial t_f=\epsilon
U_j\partial/\partial x_j$ is order one in $\epsilon$. But, when using
discrete derivatives as in (\ref{eq:filtNS2}), the same analysis
yields $\partial/\partial t_f= \epsilon U_j\delta/\delta x_j$.
Therefore, $\hat{\bar{u}}_i$ can be written:
\begin{equation}
\hat{\bar u}_i(\bk,t_f,t_s)=\hat{\bar u}_i(\bk,t_s)e^{-i\epsilon \bU\cdot\tilde\bk
t_f} + \mathcal{O}(\epsilon) ,
\label{eq:meanU}
\end{equation}
where $\hat{\bar u}_i(\bk,t_s)$ is simply the fast time average of
$\hat{\bar u}_i$, which varies in slow time due to the nonlinear
turbulent processes. The
instantaneous energy transfer rate for this skew-symmetric form as used
here (\ref{eq:Tskew}) is then
\begin{equation}
	T_{\rm skew}(\bk,t_f,t_s) = -\frac{1}{2}{\rm
	Im} \left\{ \sum_{\bk'} (\widetilde{\kappa'_\ell} + \tilde{\kappa}_\ell) \hat{\bar{u}}_k(\bk,t_s) \hat{\bar{u}}^*_k(\bk',t_s) \hat{\bar{u}}^*_\ell
	(\bk-\bk',t_s)
        e^{i \epsilon
    \mathbf{U}\cdot(\tilde{\bk'}-\tilde{\bk}-\widetilde{(\bk'-\bk)})t_f}\right\}+\mathcal
    O(\epsilon) .
\label{eq:multiscaleT}
\end{equation}
Since the turbulence is assumed to be stationary and ergodic, the
expected value $\langle T(\bk)\rangle$ is time independent and
can be estimated as a time average as follows:
\begin{equation}
	\langle T_{\rm skew}(\bk)\rangle = -\frac{1}{2}{\rm Im} \left\{ \sum_{\bk'}
		(\widetilde{\kappa'_\ell} + \tilde{\kappa}_\ell )  \langle\hat{\bar{u}}_k(\bk,t_s) \hat{\bar{u}}^*_k(\bk',t_s) \hat{\bar{u}}^*_\ell
	(\bk-\bk',t_s)\rangle
        \left\langle e^{i \epsilon
        \mathbf{U}\cdot(\tilde{\bk'}-\tilde{\bk}-\widetilde{(\bk'-\bk)})t_f}\right\rangle_{t_f}\right\}
	+ \mathcal{O}(\epsilon)
	,
\label{eq:meanT}
\end{equation}
where $\langle\cdot\rangle_{t_f}$ is the fast time average, and the
(slow) time average of the $\hat{\bar u}_i$ triple product has been
replaced by the expected value by ergodicity.

Clearly, the fast time average in (\ref{eq:meanT}) is zero unless
\begin{equation}
\omega_T=\mathbf{U}\cdot(\tilde{\bk'}-\tilde{\bk}-\widetilde{(\bk'-\bk)})=0,
\label{eq:exponent}
\end{equation}
in which case it is one. When using Fourier spectral numerics,
$\tilde\bk=\bk$ and (\ref{eq:exponent}) is satisfied identically for
all the triad interactions represented in (\ref{eq:meanT}). However,
for other numerical schemes, such as those analyzed in
Figure~\ref{fig:effective_wavenumbers}, $\tilde\bk$ is a nonlinear
function of $\bk$, and so (\ref{eq:exponent}) will generally not be
satisfied unless $|\bU\cdot\bk|=|\bU\cdot\bk'|$, severely limiting the
triad interactions that contribute to net energy transfer among
wavenumbers.  This occurs because the spatial Fourier modes that can
interact to transfer energy are determined by the wavenumbers $\bk$,
while the convective dynamics of those modes are determined by
$\tilde\bk$. The result is that the interacting wavemodes do not
maintain consistent phase relationships, essentially shutting down the
energy transfer and producing spectral anomalies like those shown in
Fig.~\ref{fig:results}. The inhibition of energy transfer due to phase
scrambling discussed here is similar to that caused by rapid rotation as
described, for example,
in \citep{cambon1997energy,waleffe1993inertial}.

Of course, this analysis is asymptotic for $\epsilon\rightarrow 0$. For
any finite $\epsilon$, there will be $\mathcal{O}(\epsilon)$
corrections because the phase scrambling effect of
the mean convective dispersion errors as described above will compete with the
nonlinear evolution of the Fourier modes. In this case, one would
expect that triad interactions for which $\omega_T\neq 0$ in
(\ref{eq:exponent}) would be weakened, rather than
completely excluded, depending on the magnitude of $\omega_T$. This
may be the reason the spectra in Fig.~\ref{fig:results} roll off smoothly
for wavenumbers with significant dispersion error.

\rdm{The condition (\ref{eq:exponent}) suggests that the strength of the
dispersion effect on the energy transfer at any wavenumber is determined by 
$U(\kappa-\tilde\kappa)=U\Delta\kappa$, which measures the rapidity of the
phase scrambling. Provided $U\Delta\kappa$ at some wavenumber is sufficiently small
compared to the rate of other processes, one would expect the
dispersion effects on the energy transfer at that wavenumber to be
negligible. This is supported by the observation that, for the third-, fourth-
and seventh-order B-splines \footnote{The lower order approximations
do not have a wide enough range of non-dispersive scales for the
scaling to hold.},
the value of $\Delta\kappa$ at the
wavenumbers where the transfer spectrum crosses that for spectral
numerics in Figure~\ref{fig:T50}  is
approximately the same ($\sim 0.4$). In addition, in simulations with
the convection velocity increased (decreased) by a factor of two (not
shown), $\Delta\kappa$ at this cross-over is decreased (increased) by
about a factor of two.}

While the above analysis was performed for the skew-symmetric form of
the nonlinear terms, the structure of the transport spectrum is
similar for other forms (see Appendix~\ref{sec:A2}), and the same
analysis applies. This suggests that the same dispersion effects
should occur for the convective, conservative, and rotational form of
the nonlinear terms, and this was confirmed numerically. Furthermore,
as is well known, the convective and conservative forms do not
necessarily conserve energy in the presence of discretization
error. 
Like the mean convective dispersion, this failure of energy
conservation is driven by the fact that
$\tilde{\bk'} - \tilde{\bk}- \widetilde{(\bk' - \bk)} \ne 0$
(Appendix~\ref{sec:A2}), which distorts the energy spectrum even when $U=0$. 
However, when $U\gg u'$, dispersion due to mean convection dominates, and the results are similar to those
presented here for the skew-symmetric form.

\section*{Discussion}
In many LES applications, the turbulence is convected with a velocity large
compared to the turbulence fluctuations (e.g., turbulent boundary layers
\citep{coles1956law}, flow through a wind turbine \citep{porte2020wind}).  In
LES of such turbulent flows, numerical dispersion error causes a decoherence of
the phase relationship among interacting Fourier modes, which results in a
reduction of the energy transfer rate from large to small resolved scales in the
direction of convection. This leads to nonphysical changes in the energy
distribution across all resolved scales.

On the other hand, the $U=0$ results indicate that the nonlinear
dispersion error has little effect on the LES spectra, despite an
inaccurate representation of the resolved scale dynamics (see
Figures~\ref{fig:E0} and~\ref{fig:T0}). 
\rdm{This is interesting because one might expect the energy transfer to be affected by phase
scrambling due to convection of the small scales by the large
scales even when $U=0$. By analogy with the scaling with the mean convection velocity,
the strength of this effect should scale with $u'$ so that, provided
$u'\Delta\kappa$ is sufficiently small compared to the rate of other
processes, the impact of dispersion on energy transfer should be
negligible. Presumably this is the case for all scales in the $U=0$ simulations
shown in Figure~\ref{fig:results}.}

However, the good $U=0$ results presented here should be interpreted with
caution, since the highly dispersive scales can have a damaging
effect in more complex
flows
\citep{ghosal1996analysis,kravchenko1997effect,chow2003further,yalla2021effects}.
Consequently, in an LES, we generally cannot expect the scales with
significant dispersion error to be dynamically meaningful.

There are two potential approaches to addressing the consequences of
dispersion error in LES. First, one could introduce a model for
$\cC_i^m$ in equation~(\ref{eq:filtNS2}) to correct for the dispersion
effects. However, it is not clear that such a model can be formulated.
\rdm{Alternatively, one could define the large scales being simulated to
only include those with sufficiently small dispersion
errors. However, the standard for
sufficiently small dispersion error depends on the convection
velocity and possibly other flow characteristics,
not just the characteristics of the derivative approximation.}

\rdm{In one approach, the large scales to be simulated can be defined} using an explicit
filter, acting in addition to the implicit filter defined by the
numerical discretization, to ensure that the scales with significant
dispersion error are not energized
\citep{lund2003use,carati2001modelling,gullbrand2003effect,winckelmans2001explicit,germano1986differential,van1995family,vasilyev1998general,marsden2002construction,bose2010grid,bose2011explicitly},
\rdm{and this was found to be effective.}
But such explicit filters are commonly not used so as to maximize
the range of scales being represented. \rdm{In this case, when
interpreting the results one should still discount
the overly dispersive scales since their dynamics are not
reliable.
Further, as shown here, the large scales are
affected negatively by interaction with the erroneous highly
dispersive scales, though perhaps this could be addressed through
refinements to the subgrid model. 
Whether explicit filters are used or not, the effective range of
dynamically resolved scales is not defined by the grid spacing, but
rather the dispersive properties of the numerical approximation to the
convection term.}

\appendix
\def\bA{\mathbf{A}}
\def\bH{\mathbf{H}}
\def\bu{\mathbf{u}}

\section{Numerical representation of the vorticity-velocity formulation}
\label{sec:A1}
The vorticity-velocity formulation introduced by
\citet{kim1987turbulence} (referred to as KMM below) is a convenient
way to solve the filtered or unfiltered Navier-Stokes equations when boundary
conditions in two spatial directions (say $x_1$ and $x_3$) are
periodic, and the numerical resolution in those directions is uniform.
However, there is a subtlety to the formulation that \gry{arises} when the
discrete second derivative operator is not equivalent to the discrete first
derivative applied twice.

In the KMM formulation, the curl and the double curl operators are
applied to the momentum equations, to obtain equations for the
vorticity and the Laplacian of the velocity. The two-component is then
solved for, and a complete representation of the velocity is obtained
from continuity. This formulation relies on three identities from vector
calculus, which must also be satisfied by the discrete operators. Let
$\tilde\nabla\cdot$, $\tilde\nabla$, $\tilde\nabla\times$, and
$\tilde\Delta$ be the discrete divergence, gradient, curl and
Laplacian operators, respectively. To recover the property of the KMM formulation
that the pressure is eliminated, we must have
\begin{equation}
	\tilde\nabla\times\tilde\nabla \psi=0 
\end{equation}
for any scalar field $\psi$. Further, to obtain the simple form used in KMM for
the double curl of the momentum equation, and to reconstruct the full velocity,
\begin{equation}
  \tilde\nabla\times\tilde\nabla\times
  \bA=-\tilde\nabla\cdot\tilde\nabla\bA+\tilde\nabla\tilde\nabla\cdot\bA
\end{equation}
must hold for any vector field $\bA$, which ensures that the second
term on the right will be zero when $\tilde\nabla\cdot\bA=0$. Both
these discrete identities hold provided the same one-dimensional
discrete derivatives are used to define the discrete divergence,
gradient and curl operators.  Finally, in KMM the 2-component of the
double curl of the momentum equation yields an equation for $\phi$,
defined as the Laplacian of $u_2$, which requires that the discrete
Laplacian obey $\tilde\Delta u_2=\tilde\nabla\cdot\tilde\nabla u_2$,
which is not generally true. So, instead, we define
$\phi=\tilde\nabla\cdot\tilde\nabla u_2$.

Consider the discrete momentum and continuity equations:
\begin{align}
  \frac{\partial \bu}{\partial t}&=-\tilde\nabla p +\bH\label{eq:FNS}\\
  \tilde\nabla\cdot\bu&=0\label{eq:FCont}
\end{align}
which could be the Navier-Stokes equations (e.g., for a DNS) or the
filtered Navier-Stokes equations (for an LES), in which case $\bu$ is
the filtered velocity. The $\bH$ term includes the
nonlinear, viscous, and model (for LES) terms. Let
$\tilde\nabla_p\cdot$ and $\tilde\nabla_p$ be the divergence and
gradient operators restricted to the (1,3) plane, and let
$\omega_2=(\tilde\nabla\times\bu)_2$. Then the discrete version of the KMM
formulation is given by
\begin{align}
  \frac{\partial\phi}{\partial
    t}&=-\tilde\nabla\times\tilde\nabla\times\bH\label{eq:phi-eq}\\
  \frac{\partial\omega_2}{\partial t}&=\tilde\nabla\times\bH\label{eq:omega-eq}\\
  \tilde\nabla\cdot\tilde\nabla u_2 &= \phi\label{eq:v-eq}\\
  \tilde\nabla_p\cdot\tilde\nabla_p u_1&=\left(\frac{\delta \omega_2}{\delta x_3} - \frac{\delta}{\delta
		x_1}\frac{\delta u_2}{\delta y} \right)\label{eq:u-eq}\\
  \tilde\nabla_p\cdot\tilde\nabla_p u_3&=\left(\frac{\delta \omega_2}{\delta x_1} - \frac{\delta}{\delta
		x_3}\frac{\delta u_2}{\delta y} \right)\label{eq:w-eq}
		.
\end{align}
With periodic boundary conditions and uniform resolution in the $x_1$
and $x_3$ directions, the discrete derivative operators in those
directions are circulant matrices, so that given $\phi$ and
$\omega_2$, (\ref{eq:v-eq})-(\ref{eq:w-eq}) can be easily solved using
discrete Fourier transforms, which is what makes the KMM formulation
efficient. This also allows one to show that the solution for $\bu$
does indeed satisfy $\tilde\nabla\cdot\bu=0$. The operators
$\tilde\nabla\cdot\tilde\nabla$ and
$\tilde\nabla_p\cdot\tilde\nabla_p$ that must be solved to recover the
velocities using (\ref{eq:v-eq})-(\ref{eq:w-eq}) are in general rank
deficient because eigenvalues associated with the Nyquist modes are
zero. This is the property that leads to checkerboard instabilities in
projection methods \citep{harlow1965numerical}. Here, the resulting singularity of the
equations is resolved by insisting that all the Nyquist modes are
zero, consistent with the Fourier cutoff filters used in the
LES. Finally, note that for spatial directions with periodic
boundary conditions and uniform resolution, there is generally no
motivation to use other than Fourier spectral
representations. Nonspectral methods are used here only to allow the
impacts of dispersion errors to be assessed, since often an LES must
be conducted for boundary conditions and resolutions for which Fourier
spectral methods are not practical.

The modified KMM formulation described here is designed to ensure that
the solutions obtained satisfy the discrete filtered or unfiltered
Navier-Stokes equations (\ref{eq:FNS})-(\ref{eq:FCont}). This is
important in the current study because we are interested in the
effects of dispersion error in these equations, which are usually solved in
practical simulations. However, it is also possible to derive the $\phi$
and $\omega_2$ equations from the momentum and mass conservation
equations before discretization, and then discretize
them \citep{kim1987turbulence,lee2015direct}. In this case, the
results are still numerical approximations to solutions of the
conservation equations, but the numerical errors are different from
those obtained by solving the discrete conservation equations.

\section{Energy transfer rate for different forms of the nonlinear terms}
\label{sec:A2}
Four different forms of the nonlinear terms are
commonly used for numerical discretization of the filtered or
unfiltered Navier-Stokes
equations. While they are equivalent analytically, they are not equivalent in the presence of
discretization error, and so result in different forms of the discrete
energy transfer spectra. These forms and associated transfer spectra
are listed below, with $\delta_j\equiv\frac{\delta}{\delta x_j}$, and
$u_i$ representing the filtered or unfiltered velocity, depending on
whether the equations being solved are filtered.
\begin{enumerate}
  \item Conservative form, $\delta_j(u_ju_i)$:
\begin{equation}
	T_{\rm cons}(\bk,t) = -{\rm Im} \left\{ \sum_{\bk'} 
		\tilde{\kappa}_\ell \hat{u}_k(\bk,t) \hat{u}^*_k(\bk',t) \hat{u}^*_\ell
	(\bk-\bk',t)  \right\}
	\label{eq:Tcons}
\end{equation}
	\item Convective form, $u_j \delta_j(u_i)$:
\begin{equation}
	T_{\rm conv}(\bk,t) = - {\rm Im}
	\left\{ \sum_{\bk'} \widetilde{\kappa'_\ell} 
		\hat{u}_k(\bk,t) \hat{u}^*_k(\bk',t) \hat{u}^*_\ell
	(\bk-\bk',t)  \right\}
	\label{eq:Tconv}
\end{equation}
	\item Skew-symmetric form, $\frac{1}{2} \left( \delta_j (u_ju_i) + u_j \delta_j (u_i) \right)$:
\begin{equation}
	T_{\rm skew}(\bk,t) = -\frac{1}{2}{\rm Im} \left\{ \sum_{\bk'} 
		(\widetilde{\kappa'_\ell} + \tilde{\kappa}_\ell) \hat{u}_k(\bk,t) \hat{u}^*_k(\bk',t) \hat{u}^*_\ell
	(\bk-\bk',t)  \right\}
	\label{eq:Tskew1}
\end{equation}
	\item Rotational form, $u_j\delta_j(u_i) - u_j \delta_i(u_j) + \frac{1}{2} \delta_i (u_j u_j)$:
\begin{equation}
	\begin{split}
	T_{\rm rot}(\bk,t) &= -{\rm Im} \left\{\sum_{\bk'}
        \tilde{\kappa'_\ell} \hat{u}_k(\bk,t) \hat{u}^*_k(\bk',t) \hat{u}^*_\ell
		(\bk-\bk',t) 
        - \tilde{\kappa'_k}
		\hat{u}_k(\bk,t) \hat{u}^*_\ell(\bk',t)
	\hat{u}^*_\ell(\bk-\bk',t)\right\}
	\label{eq:Trot}
\end{split}
\end{equation}
\end{enumerate}
Since analytically the transfer spectrum $T(\bk,t)$
is responsible for transferring energy between resolved modes,
we expect $\sum_{\bk} T(\bk,t) = 0$ for all $t$, reflecting
conservation of energy. However, it is well known that in the presence
of discretization error, both the
conservative and convective forms do not satisfy this condition. To analyze this
error, let $\tau_\ell(\bk,\bk',t) = \hat{u}_k(\bk,t) \hat{u}^*_k(\bk',t)
\hat{u}^*_\ell (\bk-\bk',t)$. Then since $\tau_\ell(\bk,\bk',t) =
\tau_\ell(-\bk',-\bk,t)$,
	\begin{align}
		\sum_{\bk} T_{\rm cons}(\bk,t) &= -{\rm Im} \left\{ \sum_{\bk} \sum_{\bk'} 
		\tilde{\kappa}_\ell \tau_\ell(\bk,\bk',t)  \right\}	
		= -\frac{1}{2}{\rm Im} \left\{ \sum_{\bk} \sum_{\bk'}  
				(\tilde{\kappa}_\ell-\tilde{\kappa'_\ell}) \tau_\ell(\bk,\bk',t) 
				\right\}.
	\end{align}
	Moreover, $(\widetilde{\kappa - \kappa'})_\ell \tau_\ell(\bk,\bk',t) = 0$ by
	continuity. Therefore,
\begin{equation}
\sum_{\bk} T_{\rm cons}(\bk,t)  = 
-\frac{1}{2}{\rm Im} \left\{ \sum_{\bk} \sum_{\bk'} 
	\left(\tilde{\kappa}_\ell - \widetilde{\kappa'_\ell} -
\widetilde{(\kappa-\kappa')}_\ell \right) \hat{u}_k(\bk,t)
\hat{u}^*_k(\bk',t) \hat{u}^*_\ell (\bk-\bk',t)  \right\}.\label{eq:sumTcons}
\end{equation}
Similarly, for the convective form,
\begin{equation}
	\begin{split}
		\sum_{\bk} T_{\rm conv}(\bk,t) &=-\frac{1}{2}{\rm Im} \left\{ \sum_{\bk} \sum_{\bk'} 
	\left(\widetilde{\kappa'_\ell} - \tilde{\kappa}_\ell +
	\widetilde{(\kappa-\kappa')}_\ell \right) \hat{u}_k(\bk,t) \hat{u}^*_k(\bk',t) \hat{u}^*_\ell (\bk-\bk',t)  \right\}
	\end{split}
.
\label{eq:sumTconv}
\end{equation}
For numerical methods other than Fourier spectral, $\tilde\bk$ is a
nonlinear function of $\bk$, so $\tilde{\bk} - \widetilde{\bk'}
-\widetilde{(\bk-\bk')} \ne 0$. Therefore, the violation of
conservation of energy for the conservative and convective forms can
be directly attributed to the fact that the effective wavenumbers of
triad interacting wavemodes do not sum to zero, as was the case for
the effects of dispersion due to mean convection (see
Eq.~\ref{eq:exponent}). Notice that the errors given by (\ref{eq:sumTcons})
and (\ref{eq:sumTconv}) are equal and opposite, and so $\sum_{\bk}
T_{\rm skew}(\bk,t) = 0$ as expected. Similarly, the rotational form
is energy conserving by construction, so $\sum_{\bk}T_{\rm rot}(\bk,t)
= 0$. Finally, note that the contribution of mean convection to the
nonlinear terms is the same in all four forms, regardless of
discretization error. Therefore, the effects of mean convection are
the same and the analysis in equations
(\ref{eq:meanU})-(\ref{eq:exponent}) holds for each form.

\begin{acknowledgments}
The authors acknowledge the generous financial support from the National
Aeronautics and Space Administration (cooperative agreement number NNX15AU40A),
the National Science Foundation (project number 1904826), and the U.S.
Department of Energy, Exascale Computing Project (subcontract number
XFC-7-70022-01 from contract number DE-AC36-08GO28308 with the National
Renewable Energy Laboratory). Thanks are also due to the Texas Advanced
Computing Center at The University of Texas at Austin for providing HPC
resources that have contributed to the research results reported here.
\end{acknowledgments}

\bibliography{main}
\end{document}